%% file: 00_main.tex
\documentclass[conference]{IEEEtran}
\IEEEoverridecommandlockouts

\usepackage{cite}
\usepackage{amsmath,amssymb,amsfonts}
\usepackage{algorithmic}
\usepackage{graphicx}
\usepackage{textcomp}
\usepackage{xcolor}
\usepackage{hyperref}
\usepackage{amsmath}
\usepackage{multirow}
\def\BibTeX{{\rm B\kern-.05em{\sc i\kern-.025em b}\kern-.08em
    T\kern-.1667em\lower.7ex\hbox{E}\kern-.125emX}}
\begin{document}

\title{A Scalable Pattern Mining Workflow for Interpretable Machine Log Analysis in High-Performance Computing Environments
\\

}

\author{\IEEEauthorblockN{1\textsuperscript{st} Shilpika}
\IEEEauthorblockA{\textit{Argonne Leadership Computing Facility} \\
\textit{Argonne National Laboratory}\\
Lemont, IL, USA \\
shilpika@anl.gov}
\and
\IEEEauthorblockN{2\textsuperscript{nd} Bethany Lusch}
\IEEEauthorblockA{\textit{Argonne Leadership Computing Facility} \\
\textit{Argonne National Laboratory}\\
Lemont, IL, USA \\
blusch@anl.gov}
\and
\IEEEauthorblockN{3\textsuperscript{rd} Eric Pershey}
\IEEEauthorblockA{\textit{Argonne Leadership Computing Facility} \\
\textit{Argonne National Laboratory}\\
Lemont, IL, USA \\
pershey@anl.gov}
\and
\IEEEauthorblockN{4\textsuperscript{th} Carlo Graziani}
\IEEEauthorblockA{\textit{Argonne Leadership Computing Facility} \\
\textit{Argonne National Laboratory}\\
Lemont, IL, USA \\
cgraziani@anl.gov}
\and
\IEEEauthorblockN{5\textsuperscript{th} Venkatram Vishwanath}
\IEEEauthorblockA{\textit{Argonne Leadership Computing Facility} \\
\textit{Argonne National Laboratory}\\
Lemont, IL, USA \\
venkat@anl.gov}
\and
\IEEEauthorblockN{6\textsuperscript{th}  Michael E. Papka}
\IEEEauthorblockA{\textit{Argonne Leadership Computing Facility} \\
\textit{Argonne National Laboratory}\\
Lemont, IL, USA \\
papka@anl.gov}
\IEEEauthorblockA{\textit{Department of Computer Science} \\
\textit{University of Illinois Chicago}\\
Chicago, IL, USA \\
papka@uic.edu}

}

\maketitle

\begin{abstract}
Modern supercomputers housed in High Performance Computing (HPC) environments generate massive volumes of log data daily, revealing intricate information and performance metrics about these complex systems. The sheer size and heterogeneous nature of HPC logs, especially text data, pose significant challenges for traditional analytical techniques. Consequently, more complex workflows are necessary for pattern extraction when analyzing these logs, enabling the discovery of underlying patterns and anomalies that may indicate system faults and help predict future failures and inefficiencies.
Our log analysis workflow investigates a combination of advanced pattern-matching and mining techniques applied to HPC log analysis. By systematically identifying frequent log patterns and pattern sequences in log messages and storing them in a finite-state automaton, such as the Aho-Corasick automaton, our workflow enables automated detection of frequent errors and fault events. To extract these patterns and sequences, we leverage information about system hierarchy and message priority. We then correlate and cluster the identified error sequences with job logs, revealing groups of applications with similar or dissimilar error signatures. This approach yields insights that inform improvements and guide real-time monitoring efforts. Our research establishes that pattern mining is vital for unlocking the full potential of log data by enabling real-time analysis and contributing to more resilient, scalable HPC systems. We demonstrate the effectiveness of our approach through summary statistics and a case study on an exascale-class system supercomputer.
\end{abstract}


\input{01_introduction}

\input{02_background}
\input{03_methodology}

\input{04_case_study_1}

\input{05_case_study_2}
\input{06_discussion}
\input{07_conclusion}

\section{Acknowledgments}
This research used resources of the Argonne Leadership Computing Facility, which is a U.S. Department of Energy Office of Science User Facility operated under contract DE-AC02-06CH11357. All authors were supported by the Office of Science, U.S. Department of Energy, under contract DE-AC02-06CH11357.

\bibliographystyle{IEEEtran}
\bibliography{00_reference}

\end{document}

%% file: 01_introduction.tex
\section{Introduction}
Log analysis in high-performance computing (HPC) systems presents a distinctive research challenge because modern supercomputers generate massive volumes of telemetry and event data across hardware, system software, runtimes, schedulers, storage, interconnects, and user applications, making the problem simultaneously one of scale, heterogeneity, and systems interpretation. Park \textit{et al.} argue that HPC log processing has become a big-data problem, not only because the data are voluminous and largely unstructured, but also because meaningful analysis requires knowledge of hardware and software behavior across multiple layers of the system stack~\cite{8049013}. These difficulties become even more severe at the leadership scale. Karimi \textit{et al.} describe parsing more than 600 million production logs from the Frontier supercomputer over a four-week period and emphasize that such logs arise from diverse runtime, hardware, and software layers with inconsistent formats, thereby complicating structure extraction, pattern discovery, and downstream mining~\cite{karimi2026instruction}. A further challenge is that failures and anomalies in HPC environments are rarely isolated; instead, they often exhibit temporal and inter-event dependencies, which makes root-cause analysis dependent on robust event-correlation techniques rather than simple message counting or isolated inspection~\cite{ren2010logmaster}. In addition, many state-of-the-art anomaly-detection methods struggle in HPC settings because operational logs frequently exhibit irregular and ambiguous time-based sequences, reducing the effectiveness of methods originally developed for more regular enterprise logs~\cite{egersdoerfer2023clusterlog}. Finally, even when logs are available for study, privacy constraints and anonymization may reduce their usefulness for failure diagnosis, creating a tension between data protection and analytic fidelity~\cite{ghiasvand2018assessing}. Overall, the central challenge in HPC log analysis is not merely storing or parsing large datasets, but transforming noisy, heterogeneous, multi-layer operational traces into timely and actionable insight for anomaly detection, failure diagnosis, and system optimization.
Log data can be broadly categorized into text-based, event-based, and numerical data being collected from multiple sources in a supercomputer. Analyzing each data type presents distinct challenges and requires tailored solutions that account for its specific structure, representation, and processing requirements. In this work, we focus our efforts on analyzing text-based log data. To mitigate and overcome some of the challenges pertaining to analyzing text-based logs, especially in high-performance computing systems, we have built an end-to-end workflow that not only processes and stores raw log representations but also helps predict the anomalies.

The contributions of this work are listed below:
\begin{itemize}
    \item[\bf 1.] Retrospective and real-time analysis of raw text-based logs.
    \item[\bf 2.] Analysis of raw text-based logs by extracting \textit{patterns} from raw log messages and storing them in an Aho-Corasick automaton.
    \item[\bf 3.] Two-level sequence mining workflow, first level to extract sequences at the intra-node process level (or thread level) and the subsequent second level to extract sequences at the job or node level. Here, a job refers to a user application running on the supercomputer.
    \item[\bf 4.] Storage of sequences in a second Aho-Corasick automaton. The automaton can be queried for searches. The queries include subsequences that can be autocompleted using a trie data structure in the automaton implementation.
\end{itemize}

The remainder of this paper is organized as follows. The related work section reviews the key concepts, prior work, and theoretical foundations relevant to the study. The methodology, the experiments, and the evaluation and discussion sections then describe the proposed approach, experimental setup, implementation details, and the criteria used to assess performance. Finally, the conclusion summarizes the main findings, discusses their implications, and outlines potential directions for future research.

%% file: 02_background.tex
\section{Related Work}
Significant efforts have been made in the past to improve HPC system resilience through log analysis. A first line of work focuses on parsing and mining unstructured HPC logs with modern language models. Karimi et al.~\cite{karimi2026instruction} present an instruction-tuned large language model (LLM) for parsing and mining heterogeneous logs from leadership-class systems, demonstrating large-scale analysis on Frontier. Closely related, Karimi et al.~\cite{karimi2025epic} introduce EPIC, a generative-AI platform for operational data analytics over multimodal HPC data, including text logs, images, and tabular records. Our work focuses on analyzing text-based data, and the resulting compressed message patterns extracted can serve as input to pipelines that employ large language models.

A second line of work studies failure prediction and job-outcome prediction from log data. Alharthi et al.~\cite{alharthi2022clairvoyant} propose \emph{Clairvoyant}, a transformer-decoder approach for node-failure prediction from system logs in large-scale systems. They extend this direction with \emph{Time Machine}, a generative model for real-time failure and lead-time prediction in HPC environments~\cite{alharthi2023timemachine}. Park et al.~\cite{park2024jobfailures} analyze production scheduler logs to study and predict job failures, while Brown et al.~\cite{brown2025xc40failures} provide a longitudinal analysis of failures on the Theta Cray XC40 supercomputer. 
Zhang et al.~\cite{zhang2021sentilog} propose \emph{SentiLog}, which applies sentiment-inspired analysis to parallel file system logs for anomaly detection. Egersdoerfer et al.~\cite{egersdoerfer2022modulelog} then introduce \emph{ModuleLog}, which organizes log events by source-code module to improve downstream anomaly detection, and subsequently propose \emph{ClusterLog}, which clusters semantically and sentimentally similar log keys to reduce irregularity and ambiguity in temporal sequences before anomaly detection~\cite{egersdoerfer2023clusterlog}.
Our workflow incorporates a mechanism, implemented via the automaton trie data structure, that allows autocompletion of an identified chain of patterns.

A third and complementary direction is the multifidelity, multiscale visual analytics path led by Shilpika et al. Rather than focusing only on text logs, this work integrates hardware logs, job logs, and environment logs collected at different temporal resolutions. The works begin with \emph{MELA}, a visual analytics tool for studying multifidelity HPC system logs~\cite{shilpika2019mela}. It is extended by a multi-level, multi-scale framework that uses multiresolution dynamic mode decomposition (mrDMD) to reveal spatial-temporal patterns across heterogeneous HPC logs~\cite{shilpika2024multilevel}. The most recent installment introduces an incremental mrDMD formulation to support faster analysis of continuously collected multifidelity system data~\cite{shilpika2024incremental}. Although these works are not restricted to text-only logs, they are highly relevant because they address cross-source correlation and multiscale interpretation of operational log data in production supercomputers, which our work addresses, although for log analysis in text format. 

Finally, Spell~\cite{7837916} and Drain~\cite{8029742} among others are log parsing frameworks that are effective in controlled settings but rely on heuristics and manual pattern design, making them fragile in the face of dynamic, evolving templates and domain-specific variations. Our workflow can accommodate evolving patterns and templates and relies little on manual intervention.
Taken together, these studies show that contemporary HPC log-analysis research is moving in three complementary directions: (i) LLM-based parsing and mining of unstructured logs, (ii) prediction and anomaly detection from system to scheduler logs, and (iii) multifidelity, multiscale analysis that correlates textual and operational logs across subsystems.

\begin{figure*}[t]
    \centering
    \includegraphics[width=\linewidth]{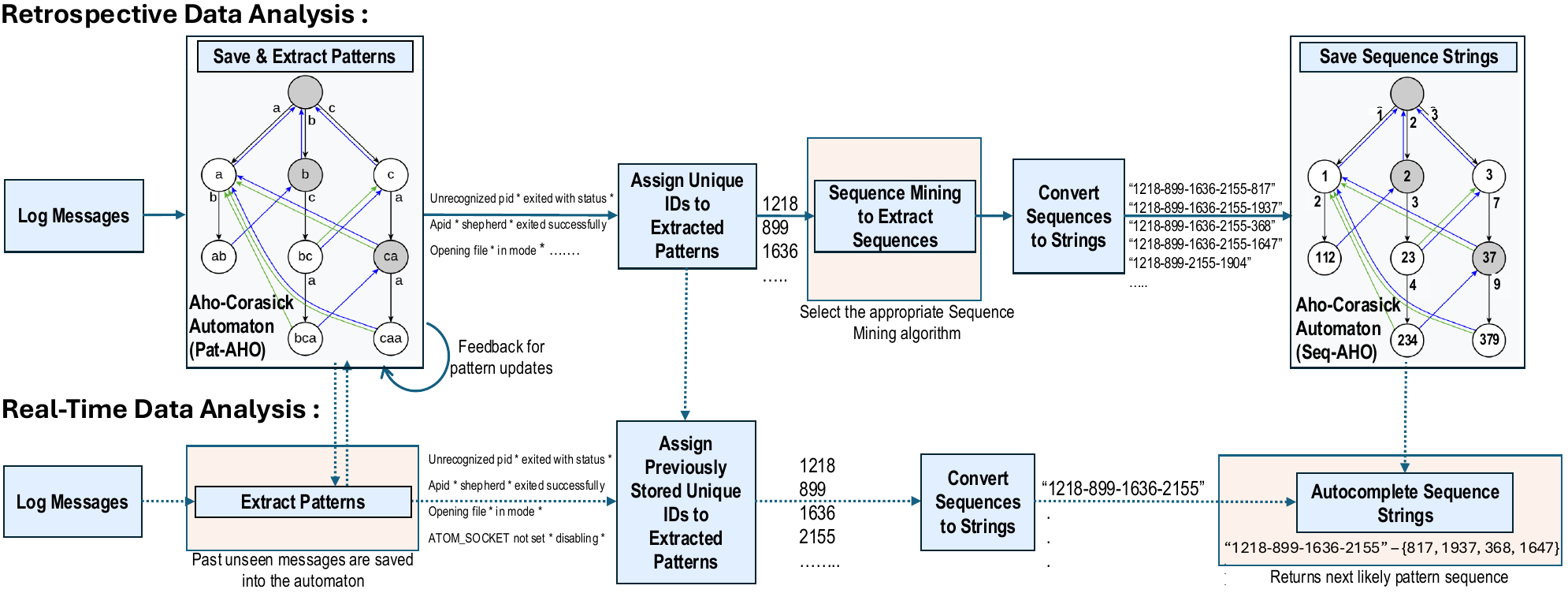}
    \caption{Log Pattern Mining Workflow Overview}
    \label{fig:archi}
\end{figure*}

\subsection{Pattern Extraction and Sequence Mining}
We use Aho–Corasick automata~\cite{ahoc,ahocWIKI} for the extraction and storage of patterns and sequences, as it has shown to be an effective mechanism in the past. 

Past works that have employed the Aho-Corasick automaton for log analysis include:
Halas (2014) studies algorithms, including Aho–Corasick, for matching log records and builds a log-matching tool~\cite{halas2014gomatch}.
Bellekens et al. (2014) present GLoP, a GPU library for security log analysis with GPU-accelerated Aho–Corasick~\cite{bellekens2014glop}.
Potharaju et al. (2013) use Aho–Corasick inside NetSieve to compute frequencies of repeated phrases extracted from network trouble tickets~\cite{potharaju2013netsieve}.
Konchagin et al. (2018) apply an Aho–Corasick-based method to search sub-traces in event logs for process mining, extending it to simultaneous search across several traces~\cite{konchagin2018processmining}.
Wang et al. (2017) use Aho–Corasick to locate dictionary phrases and compute phrase frequencies in historical IT service tickets, then use those phrases for knowledge-base construction and incident inference~\cite{wang2017itsm}.
Tuck et al. (2004) modify Aho–Corasick for intrusion detection, focusing on memory-efficient matching in Snort-style IDS pipelines~\cite{tuck2004ids}.
Tan and Sherwood (2005) describe a high-throughput Aho–Corasick-based architecture for intrusion detection and prevention~\cite{tan2005isca}.
Dimopoulos et al. (2007) present a memory-efficient reconfigurable Aho–Corasick for intrusion detection systems such as Snort~\cite{dimopoulos2007splitac}. In our work, we employ an Aho-Corasick algorithm~\cite{ahoc,ahocWIKI} in conjunction with sequence mining to efficiently store and extract log message patterns and sequences, and to predict sequences through autocompletion of these log message sequences.

Our work also uses sequential pattern mining algorithms like TUP~\cite{philippefourniervigerExampleMining,wan14830} and CLH-Miner~\cite{10.1007/978-3-030-55789-8_73} to group co-occuring patterns. In the past, sequence mining was applied to complex \textit{event} sequences~\cite{10.1016/j.knosys.2019.105241, Mavroudopoulos2021SequenceDI, 10.1155/2020/6628165, 10.1007/978-3-319-27243-6_1}, or console logs in conjunction with PCA for online problem detection~\cite{5360285}. In our work, we deal with raw text data; we first convert it into event-like patterns and then apply sequence mining algorithms to extract meaningful sequences. 

%% file: 03_methodology.tex
\section{Methodology}

~\autoref{fig:archi} covers the workflow of our analysis pipeline. There are two main processing channels: the first is retrospective analysis, and the second is real-time analysis. Our code will be made open source upon publication.

\subsection{System Overview}
During the \textbf{retrospective analysis},~\autoref{fig:archi}-top, we process historical text-based logs from multiple sources, specifically syslogs, job logs, communication logs, jobtraces, sensor logs, and application logs. The non-variable part of the log messages is retained and saved in the Aho-Corasick automaton~\cite{ahoc,ahocWIKI,pyahocorasickPyahocorasickx2014}, which we will refer to as a \texttt{pat-AHO}, since it is used to extract \textit{patterns} from raw log messages. Each of the resulting patterns is then saved in a database (RedisDB~\cite{redisRedisReference}) and assigned a unique identifier. These messages are henceforth referred to as patterns, and the identifiers are pattern IDs. Since a single log message can span multiple lines in the log data file, we then mine the co-occurring log patterns using sequence mining algorithms. This gives us an idea of which log messages are co-occurring. Henceforth, we will refer to sets of patterns that co-occur as sequences. Our analysis pipeline is flexible and can adapt to any user-specified sequence mining algorithm, including transformer-based ones. In this work, we employ sequential pattern mining, specifically TUP~\cite{philippefourniervigerExampleMining,wan14830} and CLH-Miner~\cite{10.1007/978-3-030-55789-8_73}. The results of these algorithms are a sequence of pattern IDs that the algorithm identifies as co-occurring. These sequences are then stored in a different Aho-Corasick automaton, which we will refer to as a \texttt{seq-AHO}, since it is used to save sequences of patterns. We also store the sequence occurrence counts in the seq-AHO.

During \textbf{real-time analysis},~\autoref{fig:archi}-bottom, we process log messages in real time. The logs are first stripped of the variable part, and then checked to see whether the resulting pattern was previously saved in the automaton. The search complexity of the automaton is $\mathrm{O(text\ length + number\ of\ matches)}$. Since we only look for exact matches and not substrings, the time complexity of the pattern search in the automaton depends on the size of the input (one log line) rather than the size of the automaton. Additionally, multiple log lines can be processed in parallel, leading to fast pattern lookup. If a new pattern is encountered, the pattern is registered in the automaton and RedisDB with a new pattern ID. We then look up the \texttt{seq-AHO} to determine whether the incoming stream of pattern IDs  (or sequence) exists. If the sequence is present, we can perform a lookup in the \texttt{seq-AHO} (trie data structure) to identify the next likely pattern IDs and their past occurrences count. An example is shown in the lower right of ~\autoref{fig:archi}. An incoming stream of pattern IDs  (or sequence) could be ``1218-899-1636-2155”, where each pattern ID is joined by a hyphen to form a sequence. If the \texttt{seq-AHO} has sequences with prefix ``1218-899-1636-2155” stored in the trie data structure, then a lookup could autocomplete the likely next sequence of pattern IDs,``1218-899-1636-2155” – \{817, 1937, 368, 1647\}.

\begin{figure*}[h]
    \centering
    \includegraphics[height=0.35\linewidth,width=0.85\linewidth]{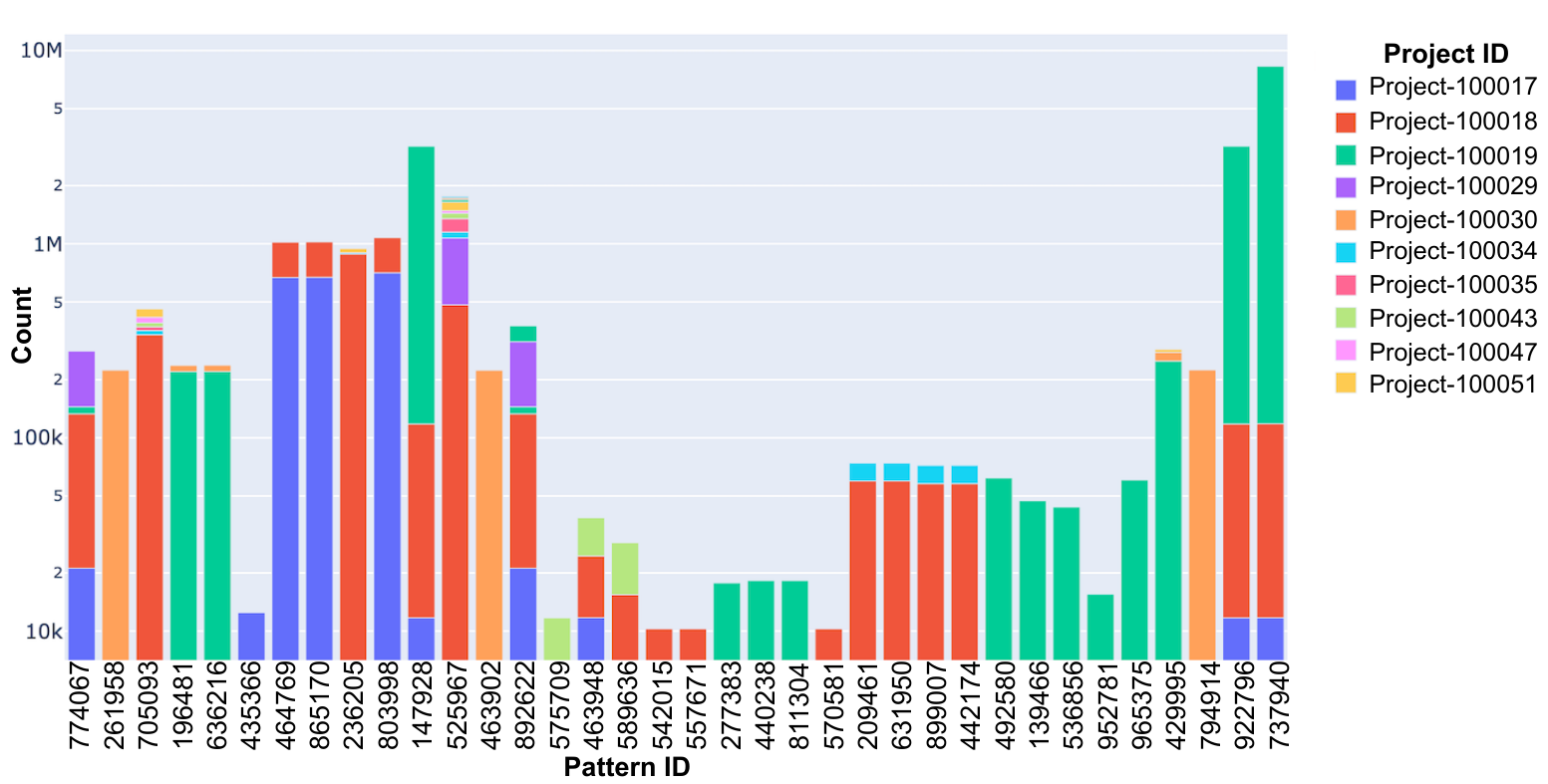}
    \caption{Most frequent log patterns grouped by projects. The x-axis shows distinct log message patterns (pattern IDs) and the y-axis shows the corresponding counts in log scale.}
    \label{fig:MP_proj_cs1}
\end{figure*}
\begin{figure*}[h]
    \centering
    \includegraphics[width=\linewidth]{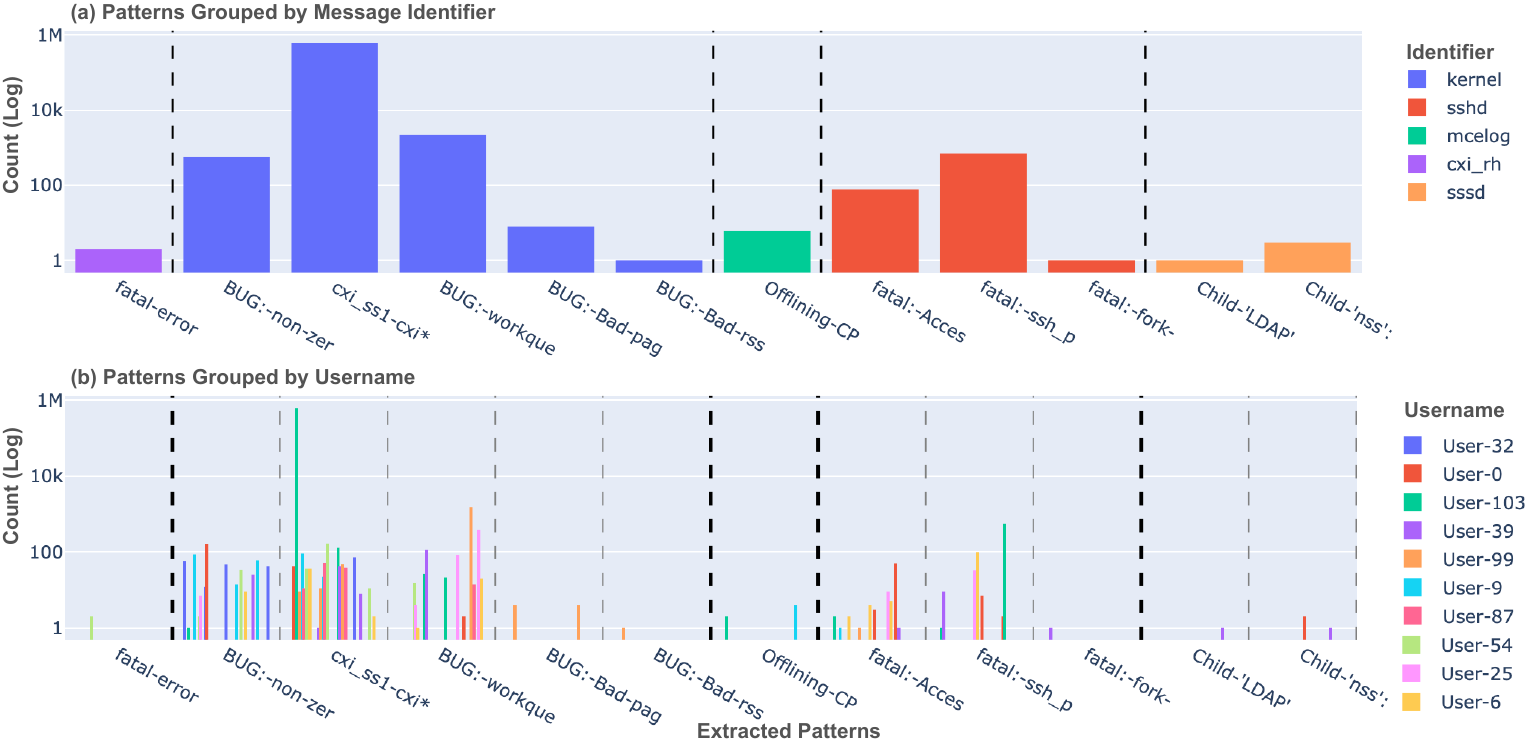}
    \caption{Most frequent log patterns grouped by identifiers (top) and users (bottom). Identifiers indicate the source component or subsystem generating a log message. The dark gray dotted vertical lines split the identifier groups, and the light gray ones split the user groups. The x-axis shows prefixes of extracted patterns.}
    \label{fig:MP_user_proj_cs1}
\end{figure*}
\begin{figure}[h]
    \centering
    \includegraphics[width=0.9\linewidth]{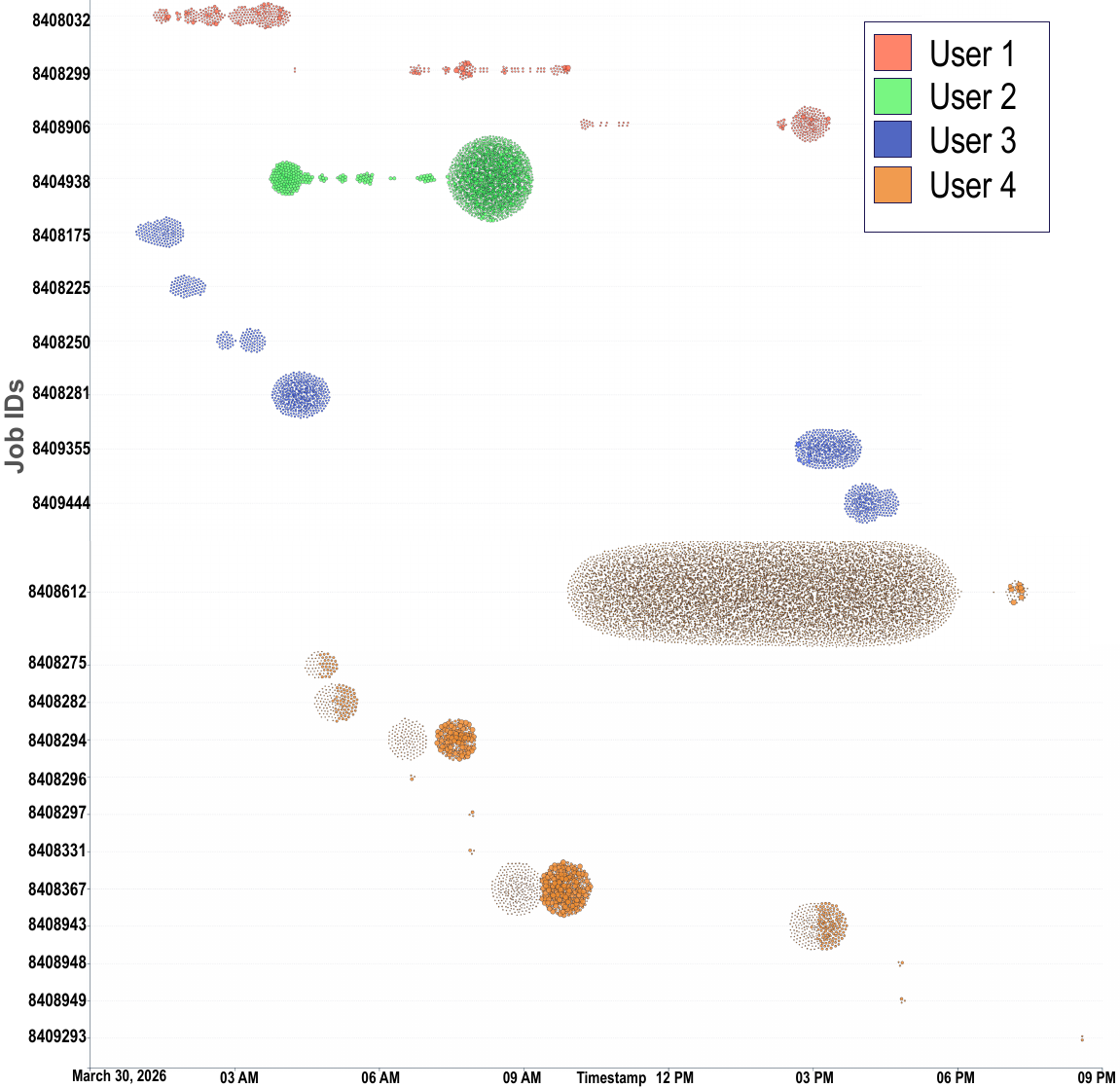}
    \caption{Error patterns in a force-directed scatterplot of jobs versus timestamps. Each point is a log pattern seen in a process on a job's node. The clusters are colored by usernames. The size of the points indicates the count of the error pattern at that timestamp.}
    \label{fig:MP_job_ts_cs1}
\end{figure}

\subsection{Sequence Mining Algorithms}
TUP~\cite{philippefourniervigerExampleMining,wan14830} is an algorithm for identifying the top K patterns with high importance. Since log lines can typically contain multiple repeated lines of informational and warning messages, and fewer lines of critical error messages, we assign an importance score either using the priority field in syslogs or manually assigning user importance scores. This weighs the fewer critical messages to have higher priority while mining patterns using TUP.

CLH-Miner~\cite{10.1007/978-3-030-55789-8_73} is an algorithm for cross-level high-importance patterns. Here, the cross-level refers to a taxonomy of patterns. The taxonomy helps classify the hierarchy of log messages, such as node-level, user-level, component-level, or job-level. The resulting group of patterns (or sequences) will contain either the nodes, users, components, or jobs that have ``seen" the sequences. Assigning the importance scores forces the algorithm to focus more on the rare critical messages. However, setting the importance scores to an equal value will treat all log message types equally. 

Section \ref{sec:exp}-(A, B) covers the data summary statistics of the pattern mining (\texttt{pat-AHO}) and the sequence mining (\texttt{pat-AHO}) part of the workflow, respectively. Additionally, Section \ref{sec:exp}-B explains how the sequence mining algorithms work in succession to extract the overall log pattern sequence.

%% file: 04_case_study_1.tex
\section{Experiments}
\label{sec:exp}
\subsection{Log Pattern Summary Statistics}

This subsection presents the summary statistics of logs from an exascale supercomputer grouped by project, job, user, and clustered trend identified from the pattern mining part of the workflow.

\subsubsection{\textbf{Project Trends}}

The \autoref{fig:MP_proj_cs1} shows a stacked bar plot summarizing the distribution of log message patterns (\texttt{pat-AHO}) of the supercomputer across projects in a day on February 2026, with the x-axis showing distinct log message patterns (pattern IDs) and the y-axis showing the corresponding counts. Each bar is stacked by project, showing how different projects contribute to the total occurrence of each pattern. The plot shows all pattern IDs whose count crosses $10,000$. The plot reveals a clear structure in the distribution. While individual projects differ in the frequency of specific patterns, groups of projects exhibit similar compositions of recurring messages. The plot also reveals patterns that are unique to a particular project for the duration of the processed log. This suggests shared operational behaviors, common software components, or comparable execution environments. Overall, the observed similarity in pattern usage indicates that log message distributions can provide meaningful signals for characterizing project-level behavior and may support downstream tasks such as workload profiling, anomaly detection, and automated grouping of related system activities.

\subsubsection{\textbf{User and Message Type Trends}}
The top plot \autoref{fig:MP_user_proj_cs1} provides a bar plot of truncated log patterns (\texttt{pat-AHO}) within the selected group of identifiers (message types), which indicates the source component or subsystem generating a log message. The bottom plot of \autoref{fig:MP_user_proj_cs1} provides a bar plot of truncated log patterns grouped by user. The dark gray dotted vertical lines split the identifier groups, and the light gray ones split the user groups. Here, we present results for January, February, and March 2026 of the supercomputer where the per-user job count is greater than $10$. This visualization shows that several users across different projects exhibit similar log message patterns, while others exhibit unique patterns. This indicates cross and unique commonalities in user activity or system behavior. These shared patterns suggest that similarities observed at the project level (Section \ref{sec:exp}-(A-1) may also extend to the user level, providing further evidence of recurring operational structures and offering useful insight for behavioral profiling, workload characterization, and anomaly analysis.

\subsubsection{\textbf{Job Timeline Trends}}
Until now, we have observed trends in projects, identifiers, and user groups filtered by frequency criteria. Here, we look more in-depth at patterns within a job. 
The temporal scatter plot, \autoref{fig:MP_job_ts_cs1}, illustrates how log messages generated by user jobs are distributed over time, with time shown on the x-axis and individual user jobs shown on the y-axis. Each point represents a log message occurrence and is colored by username, enabling comparison of temporal activity pattern frequency across users and jobs. The points that are overlapping are spread out using a force-directed algorithm. The size of the points indicates the count of the error pattern at that timestamp. The visualization reveals that jobs associated with the same user often exhibit similar temporal structures in their log message distributions. However, these patterns vary in density: for some jobs, log messages are distributed broadly across the execution timeline, whereas in others they appear more temporally concentrated. This suggests that user-specific job behavior can manifest through recurring temporal signatures, while differences in spread may reflect variation in workload phases, runtime characteristics, or system interactions.

\subsubsection{\textbf{Cluster Trends}}
\begin{figure}[h]
    \centering
    \includegraphics[width=\linewidth]{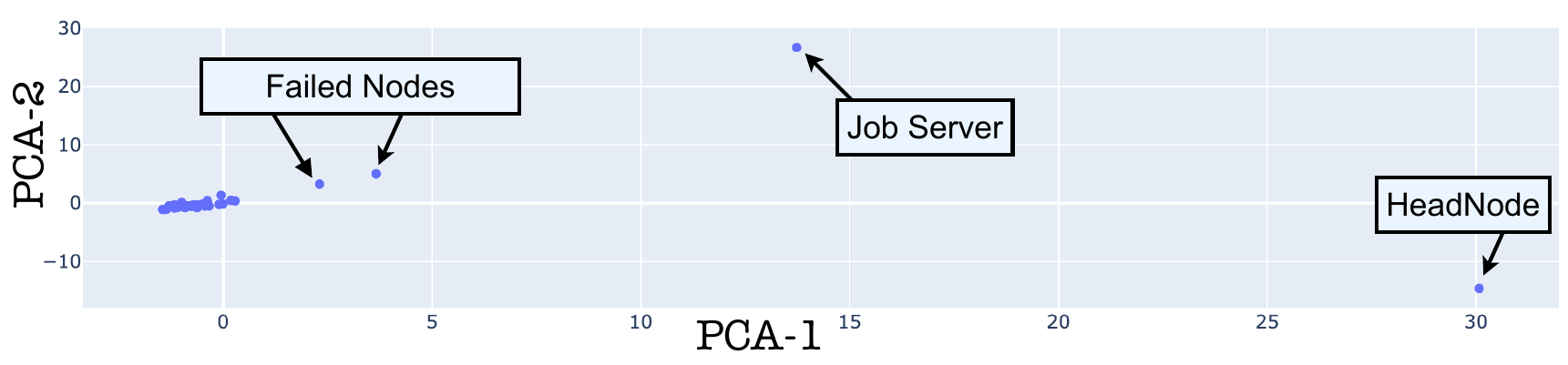}
    \caption{PCA plot of the node-level variation in log messages computed by projecting a node-by-pattern frequency matrix into a lower-dimensional space using PCA.}
    \label{fig:MP_cluster_pca_cs1}
\end{figure}
Here, we applied pattern mining (\texttt{pat-AHO}) to the scheduler, console, sensor, and other jobtrace information on logs that indicated communication failure messages.
The PCA plot, \autoref{fig:MP_cluster_pca_cs1}, summarizes node-level variation in log message behavior by projecting a node-by-pattern frequency matrix into a lower-dimensional space. In this representation, each point corresponds to a node, and the input features are the frequencies of observed log message patterns. The resulting projection shows distinct clusters corresponding to the head node, error-prone nodes, the remaining compute nodes, and the server node, indicating that different node roles and failure states produce separable log-message signatures. Notably, the separation of error nodes from the head-node cluster helps identify nodes that appear to be losing communication with the head node. These results suggest that analysis of log-pattern frequencies can support visual diagnosis of system state, role-specific behavior, and communication-related anomalies in large-scale computing environments.

%% file: 05_case_study_2.tex
\subsection{Case Study}
\label{sec:cs2}

\begin{figure*}[h]
    \centering
    \includegraphics[width=0.9\linewidth]{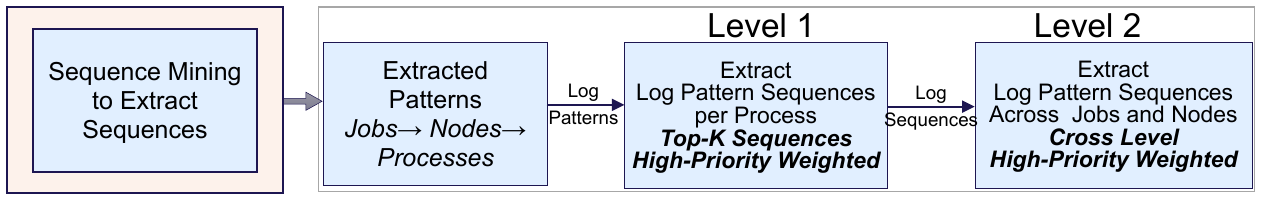}
    \caption{Overview of the sequence mining workflow.}
    \label{fig:seq_archi_cs1}
\end{figure*}

\begin{figure*}[h]
    \centering
    \includegraphics[width=0.9\linewidth]{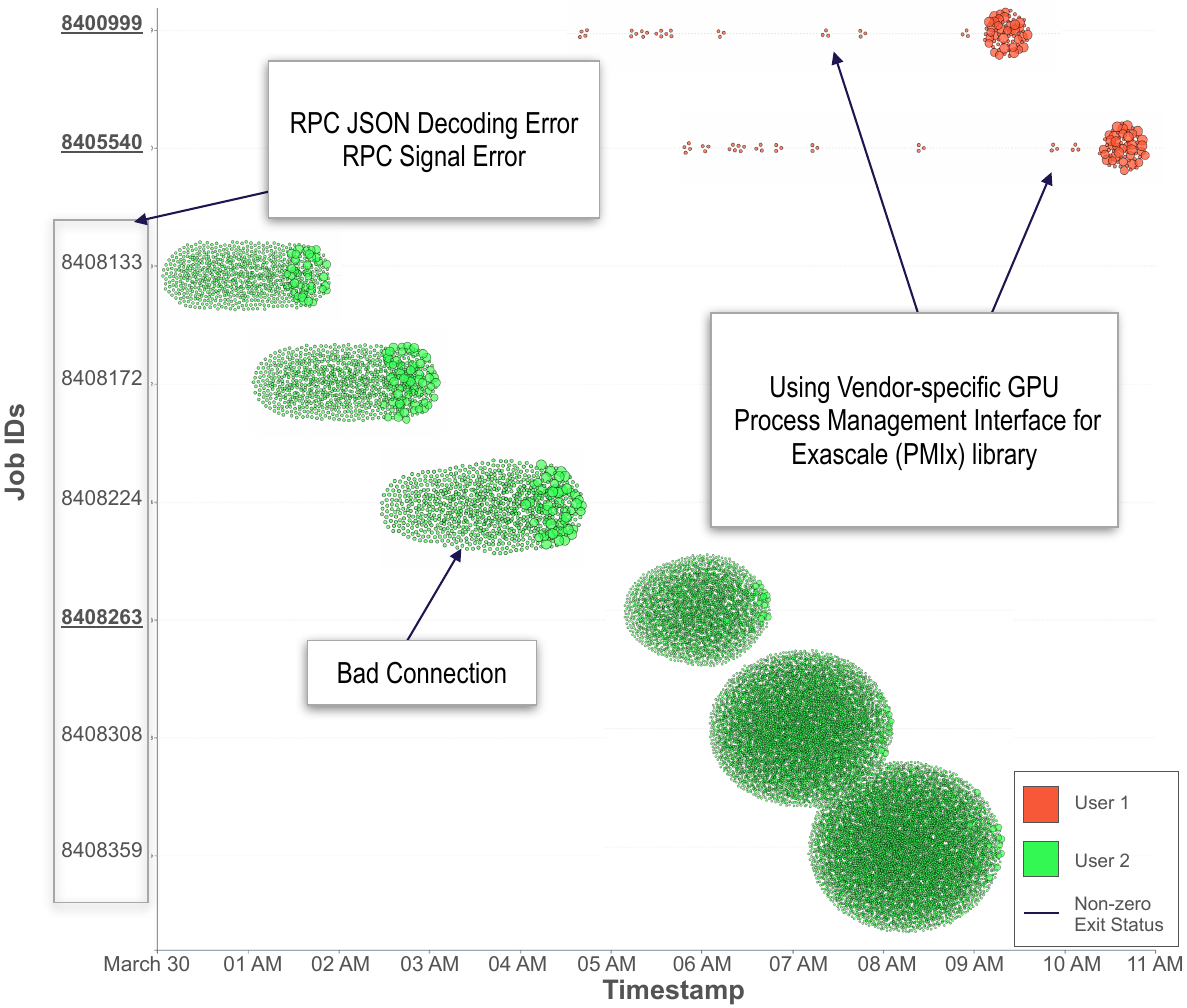}
    \caption{Error patterns in a force-directed scatterplot of jobs versus timestamps. Each point is a log pattern seen in a process of a job's node. The size of the points indicates the count of the error pattern at that timestamp. The clusters are colored by usernames. The annotations show summaries of the sequences that are identified by the sequence mining workflow in~\autoref{fig:seq_archi_cs1}. The underlined y-axis labels are jobs with non-zero exit status.}
    \label{fig:cs2}
\end{figure*}

\bigskip
\noindent
\begin{table}[h]
    \centering
    \caption{Importance Scores for Syslog Priority}
    \label{tab:sev_weight}
    \begin{tabular}{|c|c|c|} \hline
    \small Priority & Type & Importance Score   \\ \hline
    \texttt{0} & EMERGENCY &  100,000 \\ \hline
    \texttt{1} & ALERT &  90,000 \\ \hline
    \texttt{2} & CRITICAL & 70,000 \\ \hline
    \texttt{3} & ERROR &  60,000 \\ \hline
    \texttt{4} & WARNING &  8,000 \\ \hline
    \texttt{5} & NOTICE & 2,000 \\ \hline
    \texttt{6} & INFORMATIONAL &  1 \\ \hline
    \texttt{7} & DEBUG &  1 \\ \hline
    \end{tabular}
\end{table}

So far, we have seen how extracting patterns from log messages helps gain useful insights into system usage and identify trends at various levels of the system hierarchy. In this section, we will examine how extracting temporally aligned patterns can reveal more insights into the machines and their usage. We will refer to these temporally aligned patterns as sequences. A sequence is a group of patterns that co-occur within defined rules. The rules specify the maximum time window length and the number of sequences to extract within it. 

For this case study, we use syslogs of an exascale supercomputer for a subset of jobs on the 30th March 2026. Each log line corresponds to a process within a node in a user's job. Throughout the runtime of a multinode job, numerous processes on a node record messages that reflect their internal state. Therefore, to extract useful sequences relating to a process within a node, we first group the log patterns by job, node, and process. We then apply sequence mining algorithms at two levels, refer ~\autoref{fig:seq_archi_cs1}. At Level 1, we group the patterns per process within a node and extract process-specific sequences. This is an embarrassingly parallel problem to compute. Level 2 then applies the sequence-mining algorithm to the Level 1 sequences to extract common sequences across nodes or jobs. We apply two separate sequence mining algorithms at each level. Level 1 uses the “TUP algorithm,” and Level 2 uses the “CLH-Miner Algorithm”. 

\subsubsection{Level 1 Sequence Extraction}
The TUP algorithm discovers the top-k high-importance sequences in a complex event sequence that contains importance information. For this case study, we have set k to 10. The importance scores can either be user-defined or set by using the syslog message priority field. Setting the importance score to an equal number gives equal importance to all log patterns. The TUP algorithm extracts the top-k process-level sequences, weighted or prioritized by user-specific importance scores. We use the syslog message priority for syslog sequence extraction. 
\autoref{tab:sev_weight} shows the values we have set for each priority type in this study. We found that setting a larger difference in importance scores between lower- and higher-priority log messages yields more accurate results. Log messages can be duplicated, and the same messages may appear in large bursts within a small timeframe, as seen in point sizes in \autoref{fig:MP_job_ts_cs1} and \autoref{fig:cs2}. We retain only the unique messages at every 10-second interval at the process level. The importance scores are weighted by the count of occurrence and the weight specified in \autoref{tab:sev_weight}.

\subsubsection{Level 2 Sequence Extraction}

At Level 2, we first assign unique IDs to process-level sequences extracted at Level 1, refer~\autoref{fig:seq_archi_cs1}. We then group the sequences by jobs and nodes. Following this step, we apply the CLH-Miner Algorithm. The CLH-Miner algorithm discovers cross-level high-importance sequences in a dataset containing importance scores. Here, we can define a hierarchy of nodes or jobs that have seen the sequences previously extracted at Level 1. For example, if the sequence \{1,2,3\} is reported on multiple nodes, we can provide this information to the CLH-Miner, and, along with sequence grouping, the algorithm identifies nodes that have reported similar sets of sequences \{1,2,3\}--\{4,5,6\}. Note that at Level 2, we extract sequences of process-level sequences. These sequences are then stored in the \texttt{seq-AHO} automaton for retrieval and future analysis.

\autoref{fig:cs2} shows the jobs along the y-axis and the timeline along the x-axis. We have filtered 8 jobs belonging to two users running on the system on March 30th. The underlined jobs are the ones that failed with a non-zero exit code, and each failed with a user termination from the shell. Each point is a log message pattern reported by a process within a node. To avoid overlapping points, we have used a force-directed algorithm. The top three jobs of user-2 two uses lesser nodes than the bottom three jobs. We see that the bottom three jobs share a unique visual signature compared to the top three jobs and the jobs belonging to user 1. At first glance, it seems that the pattern signature may be unique across users. However, we are interested in the sequences of these patterns over time and across jobs or nodes. When we extract the process-level sequences at Level 1 and job-level sequences at Level 2. We found that job id $8408172$ has a very unique log message signature pertaining to \textit{bad network connection} reported by node \texttt{x4610c4s3b0n0}. Job Ids $8400999$ and $8405540$, belonging to user-1, have a unique signature, \textit{RPC-unknown-none-:-error-JSON-decoding-error-at-byte-of-...}, where a remote procedure client called a backend/helper program that was supposed to return JSON, but instead received an unexpected text leading to a runtime error. User-1 jobs also reported hardware vendor-specific messages. All eight jobs reported an application abort and clean-up messages, which were captured as a separate stream of sequences.
Since we set the startup and cleanup messages as low-importance, these informational types of messages were not captured by the sequence mining algorithms. In this case study, we conclude that even though summary statistics and other visual signatures can indicate high-level trends, a deeper analysis can yield more distinct signatures. We are interested in understanding the behavior and identifying message signatures. This could eventually drive further analysis, including root-cause identification, which is not in the scope of this work.

%% file: 06_discussion.tex
\section{Evaluation and Discussion}

In this section, we evaluate the effectiveness of our workflow in extracting patterns and sequences from syslogs of an exascale supercomputer. We have analyzed syslogs for the months of January, February, and March of 2026. The total size of the dataset is roughly 900GB. In this section, we process only the syslogs pertaining to jobs on the supercomputer. We evaluated our workflow on an NVIDIA DGX A100 node with two AMD Rome CPUs and 1TB of DDR4 Memory.

\subsection{Analytical and Throughput Evaluation}

\bigskip
\noindent
\begin{table}[h]
    \centering
    \caption{Automaton Pattern and Sequence Retrieval Time}
    \label{tab:AHOpatSeqTime}
    \begin{tabular}{|p{0.16\linewidth}|p{0.1\linewidth}|c|c|p{0.1\linewidth}|} \hline
    \small Automaton & Size & Data Size & String Length & Time (ms)  \\ \hline
    \multirow{4}{4em}{\texttt{pat-AHO}}  & 8.8MB & ~900GB & 65 & 0.35 \\ 
    & 8.8MB & ~900GB & 131 & 0.33 \\ 
    & 8.8MB & ~900GB & 659 & 1.12 \\ 
    & 8.8MB & ~900GB & 989 & 1.47 \\ \hline
    \multirow{4}{4em}{\texttt{seq-AHO}}  & 12MB & ~900GB & 72 & 0.42 \\ 
     & 12MB & ~900GB & 145 & 0.63 \\ 
     & 12MB & ~900GB & 729 & 1.89 \\ 
     & 12MB & ~900GB & 1094 & 3.05 \\ \hline
    \end{tabular}
\end{table}

\bigskip
\noindent
\begin{table}[h]
    \centering
    \caption{Level 1 Sequence Retrieval: Time and Accuracy}
    \label{tab:LV1SeqTime}
\begin{tabular}{|p{0.05\linewidth}|p{0.06\linewidth}|p{0.06\linewidth}|p{0.08\linewidth}|p{0.1\linewidth}|p{0.06\linewidth}|p{0.1\linewidth}|p{0.1\linewidth}|} \hline
\small ID & Jobs & Proce-sses & Log Lines & High Priority Lines & Accu-racy & Time Per Process (s) & Total Time (s) \\ \hline
0 & 5 & 1808 & 15454 & 12956 & 100.0 & 0.3371 & 609.41\\ \hline
1 & 4 & 319 & 965 & 388 & 100.0 & 0.3348 & 106.7\\ \hline
2 & 8 & 3420 & 18648 & 13737 & 100.0 & 0.3356 & 1147.8 \\ \hline
3 & 2 & 2202 & 27767 & 26742 & 96.21 & 0.3681 & 810.4\\ \hline
4 & 2 & 1982 & 44254 & 42891 & 100.0 & 0.3366 & 667.0\\ \hline
5 & 5 & 864 & 11337 & 9800 & 100.0 & 0.3431 & 296.3\\ \hline
6 & 4 & 2282 & 281665 & 281066 & 100.0 & 0.3391 & 773.8\\ \hline
7 & 8 & 189 & 10646 & 10052 & 100.0 & 0.3303 & 62.4\\ \hline
8 & 65 & 4681 & 196624 & 182043 & 100.0 & 0.3590 & 1680.47\\ \hline
9 & 2 & 3689 & 109099 & 108543 & 100.0 & 0.3368 & 1242.3\\ \hline
\end{tabular}
\end{table}

\begin{table*}[h]
  \centering
  \caption{Level 2 Sequence Retrieval: Time and Accuracy}
    \label{tab:LV2SeqTime}
  \begin{tabular}{lllclllcc}
    \hline
    \small ID & Jobs & Processes & Log Lines & High Priority Lines & Min Importance & Accuracy & Time (s) & Total Time (s)\\ \hline
0 & 2 & 2202 & 27767 & 26742 & 10 & 100.0 & 0.57 & 2.23 \\ \hline
0 & 2 & 2202 & 27767 & 26742 & 100 & 100.0 & 0.57 & 2.60\\ \hline
0 & 2 & 2202 & 27767 & 26742 & 1000 & 100.0 & 0.57 & 2.95\\ \hline
0 & 2 & 2202 & 27767 & 26742 & 5000 & 100.0 & 0.57 & 3.31\\ \hline
1 & 4 & 2282 & 281665 & 281066 & 10 & 100.0 & 0.58 & 3.10 \\ \hline
1 & 4 & 2282 & 281665 & 281066 & 100 & 100.0 & 0.57 & 3.52 \\ \hline
1 & 4 & 2282 & 281665 & 281066 & 1000 & 100.0 & 0.58 & 3.94 \\ \hline
1 & 4 & 2282 & 281665 & 281066 & 5000 & 100.0 & 0.57 & 4.35 \\ \hline
2 & 4 & 166 & 1714 & 829 & 10 & 100.0 & 0.56 & 2.25\\ \hline
2 & 4 & 166 & 1714 & 829 & 100 & 100.0 & 0.56 & 2.58\\ \hline
2 & 4 & 166 & 1714 & 829 & 1000 & 100.0 & 0.56 & 2.92\\ \hline
2 & 4 & 166 & 1714 & 829 & 5000 & 100.0 & 0.56 & 3.25\\ \hline
3 & 5 & 1808 & 15454 & 12956 & 10 & 100.0 & 0.56 & 1.57\\ \hline
3 & 5 & 1808 & 15454 & 12956 & 100 & 100.0 & 0.58 & 1.92\\ \hline
3 & 5 & 1808 & 15454 & 12956 & 1000 & 100.0 & 0.57 & 2.26\\ \hline
3 & 5 & 1808 & 15454 & 12956 & 5000 & 100.0 & 0.56 & 2.61\\ \hline
4 & 4 & 319 & 965 & 388 & 10 & 100.0 & 0.56 & 1.32\\ \hline
4 & 4 & 319 & 965 & 388 & 100 & 100.0 & 0.56 & 1.65\\ \hline
4 & 4 & 319 & 965 & 388 & 1000 & 100.0 & 0.56 & 1.98\\ \hline
4 & 4 & 319 & 965 & 388 & 5000 & 100.0 & 0.56 & 2.32\\ \hline
5 & 2 & 1982 & 44254 & 42891 & 10 & 100.0 & 0.59 & 1.80\\ \hline
5 & 2 & 1982 & 44254 & 42891 & 100 & 100.0 & 0.58 & 2.15\\ \hline
5 & 2 & 1982 & 44254 & 42891 & 1000 & 100.0 & 0.58 & 2.50\\ \hline
5 & 2 & 1982 & 44254 & 42891 & 5000 & 100.0 & 0.57 & 2.86\\ \hline
6 & 8 & 3420 & 18648 & 13737 & 10 & 100.0 & 0.56 & 2.05\\ \hline
6 & 8 & 3420 & 18648 & 13737 & 100 & 100.0 & 0.60 & 2.40\\ \hline
6 & 8 & 3420 & 18648 & 13737 & 1000 & 100.0 & 0.57 & 2.75\\ \hline
6 & 8 & 3420 & 18648 & 13737 & 5000 & 100.0 & 0.56 & 3.10\\ \hline
7 & 5 & 864 & 11337 & 9800 & 10 & 100.0 & 0.62 & 1.68\\ \hline
7 & 5 & 864 & 11337 & 9800 & 100 & 100.0 & 0.59 & 2.06\\ \hline
7 & 5 & 864 & 11337 & 9800 & 1000 & 100.0 & 0.61 & 2.45\\ \hline
7 & 5 & 864 & 11337 & 9800 & 5000 & 100.0 & 0.59 & 2.83\\ \hline
8 & 8 & 464 & 27747 & 3345 & 10 & 100.0 & 0.67 & 3.037\\ \hline
8 & 8 & 464 & 27747 & 3345 & 100 & 100.0 & 0.69 & 3.5\\ \hline
8 & 8 & 464 & 27747 & 3345 & 1000 & 100.0 & 0.66 & 3.9\\ \hline
8 & 8 & 464 & 27747 & 3345 & 5000 & 100.0 & 0.67 & 4.3\\ \hline
9 & 2 & 3689 & 109099 & 108543 & 10 & 100.0 & 0.56 & 2.83\\ \hline
9 & 2 & 3689 & 109099 & 108543 & 100 & 100.0 & 0.56 & 2.06\\ \hline
9 & 2 & 3689 & 109099 & 108543 & 1000 & 100.0 & 0.56 & 2.46\\ \hline
9 & 2 & 3689 & 109099 & 108543 & 5000 & 100.0 & 0.57 & 2.85\\ \hline

  \end{tabular}

\end{table*}
 
~\autoref{tab:AHOpatSeqTime} shows the retrieval time for patterns (pat-AHO) and sequences (seq-AHO) stored in the automaton. Note that to store information relevant to ~900GB of log data, we only require a storage space of 8.8MB for patterns and 12MB for sequences in pickle format. The ID column is a unique file ID. We vary the input character length from 60 to 1000 characters and measure the time required to retrieve these patterns. The search complexity of the automaton is $\mathrm{O(text\ length + number\ of\ matches)}$~\cite{ahoc}. Here, we only look for exact matches as we are not interested in substring matching. However, if the user requires substring matching, it is an option they can choose. Hence, retrieval time is dependent on the text length, which is in a few milliseconds as seen in \autoref{tab:AHOpatSeqTime}. Note that the seq-AHO input is a string of pattern IDs extracted by the sequence mining Level 1 and Level 2 output (\autoref{fig:seq_archi_cs1}).

\autoref{tab:LV1SeqTime} shows the time and accuracy at Level 1 of the sequence mining part of the workflow for a subset of failed jobs in the dataset. At Level 1, we retrieve sets of co-occurring patterns. As mentioned previously, a single log message can span multiple log lines. We use level 1 to extract these temporally co-occurring log lines for a process on a node of a supercomputer. Each row of \autoref{tab:LV1SeqTime} contains a unique file ID (ID), total number of jobs (Jobs), total number of processes (Processes), total number of log lines (Log Lines), total number of high-importance (priority less than 5.0) log lines (High Priority Lines), accuracy of retrieval (Accuracy), time for retrieval of the sequences per process in seconds (Time Per Process (s)), and total time for retrieval of the sequences in seconds (Total Time (s)). Accuracy is measured by whether the algorithm retrieves all high-priority sequences. We are only looking for high-priority sequences, as they are fewer in number and more critical than warning, alert, and informational messages, which dominate the log lines. Note that our sequence mining results include informational, warning, and alert patterns in the temporal vicinity of the high-priority patterns. \autoref{tab:LV1SeqTime} shows that we can retrieve all high-priority messages with over $96\%$ accuracy and, in most cases, $100\%$ accuracy.  Note that level 1 sequence mining operates on a single process at a time, making it an embarrassingly parallel problem. We can isolate each job and processes within a job's node independently of one another, and each takes roughly 0.3 seconds of compute time.

\autoref{tab:LV2SeqTime} shows the Time and Accuracy at Level 2 of the Sequence Mining part of the workflow for a subset of failed jobs in the dataset. At Level 2, we retrieve sets of co-occurring sequences. Each of these co-occurring sequences is a process-level sequence from the output of level 1. Each row of \autoref{tab:LV2SeqTime} contains a unique file ID (ID), the total number of jobs (Jobs), total processes (Processes), total log lines (Log Lines), total high-importance (priority less than 5.0) log lines (High Priority Lines), minimum importance (Min Importance), accuracy of retrieval, and time for retrieval of the sequences (Time in seconds) and total time (Total time in seconds). The total time includes time for grouping by nodes or jobs and hierarchy creation. Minimum importance is a user-defined threshold that determines the minimum importance score for the output sequence. Using this field, we can retrieve only the relevant sequences and eliminate noisy informational message patterns.  However, if the user prefers to view more warning alerts and informational messages, then the min support value should be reduced. Accuracy is measured by whether the algorithm retrieves all high-priority sequences. Here again, we are only looking for high-priority sequences. \autoref{tab:LV2SeqTime} shows that we can retrieve all high-priority messages with $100\%$ accuracy.  Note that if level 1 missed a high-importance pattern, then that would be missed by level 2 as well, since the input to level 2 is the output of level 1. The accuracy only indicates whether level 2 sequence mining retrieves all the high-importance sequences identified by level 1. Note that level 2 sequence mining can be programmed to operate at the node level or the job level. Here, we have grouped sequences from all jobs that meet the specified minimum importance score. Since we are only looking for high-importance patterns, the level 2 part of the workflow roughly takes 0.5 seconds to compute. The total time includes time for grouping by nodes or jobs and hierarchy creation, and the total processing time is on average 2.5 seconds.

Our results highlight the value of pattern mining for real-time monitoring, fault characterization, and enhancing resilience in modern HPC systems. The compressed data, in the form of patterns and sequences, can be used as input for other types of analysis, for example, large language model training and prompting. The log size can be prohibitive for RAG-like (retrieval augmented generation) workflows. The pattern extraction also enables us to use sequence mining approaches, such as the TUP and CLH-Miner algorithms, which have previously been used for event data analysis.

As a part of the future work for this workflow, we plan to automate the setting of the minimum support variable. Currently, the drawback of this approach is that the user must make an informed decision based on their domain expertise to set the minimum support, which may not be ideal in some cases. Another drawback of this approach is that we rely on priority or importance information for the sequence mining approach, which must be set by the user for some datasets. Another part of future work is to automate the detection of message importance through a preprocessing data analytics step. We plan to extend our current workflow for root-cause analysis, which is not in the scope of this study because we are processing only text data, whereas supercomputer logs are multivariate, spanning text, events, and numerical data. For effective root-cause analysis, a holistic view with multivariate data processing pipelines is necessary.

%% file: 07_conclusion.tex
\section{Conclusion}

This work presents a scalable log analysis workflow for extracting, detecting, and interpreting fault-related patterns in large-scale HPC environments. By combining pattern mining algorithms with the Aho-Corasick finite-state automata for efficient sequence detection, the proposed approach enables systematic identification of frequent error patterns and error sequences from heterogeneous system logs. Associating these mined sequences and patterns across logs, clustering, and visualizations, further allows supercomputer jobs to be grouped by shared error signatures, providing insight into recurring failure modes, anomalous behavior, and system–application interactions. Through use cases and summary statistics on an exascale-class supercomputer, we demonstrate that the workflow can uncover meaningful error patterns and support automated analysis of large volumes of log data. These results highlight the value of pattern mining as a foundation for real-time monitoring, fault characterization, and reliability improvement in modern HPC systems. The compressed format of data in patterns can be used as input for other types of analysis, for example, AI and LLM training, and prompting. The pattern extraction also enables us to use sequence mining approaches, such as the TUP and CLH-Miner algorithms, which have previously been used for event data analysis. Our workflow enables online anomaly detection and predictive failure analysis with near 100\% accuracy for rare, high-priority messages, and integrates with operational monitoring frameworks to provide high-speed support in building more resilient, efficient supercomputing infrastructures.